\documentstyle[12pt,aasms4]{article}

\def\mh{km s$^{-1}$ Mpc$^{-1}$}
\def\apgt{$^>$ \hskip -0.4cm $_\sim$~}
\def\kms{km s$^{-1}$}
\def\secpoint{\mbox{$''\mskip-7.6mu.\,$}}
\def\eg{{\it e.g.}}
\def\etal{{\it et al.}}

\begin{document}

\title{Observations of GRB 970228 and GRB 970508, and the Neutron-Star Merger Model}

\author{{\it Kailash C.~Sahu}\footnote{Space Telescope Science Institute, 
3700 San Martin Drive, Baltimore, MD 21218},
{\it Mario Livio}$^{1}$,
{\it Larry Petro}$^{1}$,
{\it Howard E.~Bond}$^{1}$,
{\it F.~Duccio Macchetto}$^{1,}$\footnote{On assignment from the Space Science 
Department of ESA},
{\it Titus J. Galama}\footnote{Astronomical Institute ``Anton Pannekoek", 
University of Amsterdam, \& Center for High Energy Astrophysics, 
Kruislaan 403, 1098 SJ Amsterdam, The Netherlands},
{\it Paul J.~Groot}$^{3}$,
{\it Jan van Paradijs}$^{3,}$\footnote{Physics Department,
University of Alabama in Huntsville, Huntsville, AL 35899},
{\it Chryssa Kouveliotou}\footnote{Universities Space Research Association, 
NASA Marshall Space Flight Center, ES-84, Huntsville, AL 35812, USA}}
\authoremail{ksahu@stsci.edu}

\begin{abstract}
We present the discovery observations for the optical counterpart of
the $\gamma$-ray burster GRB 970508 and discuss its light curve
in the context of the fireball model. We analyze the HST data for
this object, and conclude that any underlying galaxy must be fainter
than $R$ = 25.5. We also present a detailed analysis of
the HST images of GRB 970228 claimed to show a proper motion of the optical
counterpart  and conclude that, within the uncertainties,
there is no proper motion. We examine several aspects of the
neutron-star merger model for $\gamma$-ray bursts. In particular, we use
this model to predict the redshift distribution of $\gamma$-ray bursters,
and adopting a recent determination of
the cosmic star-formation history, we show that the predicted distribution 
of $\log N - \log P$ relation is consistent with that observed for GRBs.

\end{abstract}


\section{Introduction}

Recent observations of the two $\gamma$-ray bursts (GRBs) GRB~970228
and GRB~970508 have allowed unprecedented progress in our
understanding of their sources, due to the fact that X-ray and optical
counterparts have now been identified (Costa \etal\ 1997a,~b; Heise
\etal\ 1997; Piro \etal\ 1997; van Paradijs \etal\ 1997; Sahu \etal\ 1997; 
Galama \etal\ 1997a; Bond 1997; Djorgovski \etal\ 1997a; and
references therein).  In particular, the discovery of an extended
source which {\it may be} the host galaxy of GRB~970228 (van Paradijs
\etal\ 1997; Sahu \etal\ 1997), and the detection of an absorption-
and emission-line system at $z = 0.835$ in the optical spectrum of
GRB~970508 (Metzger \etal\ 1997a,~b), which may arise from a host (or
intervening) galaxy, provide (if the identification of the
sources is correct) the first direct evidence  that GRBs are
at cosmological distances. 

However, there does remain one serious  difficulty with the cosmological
hypothesis, namely, the claim that a significant proper motion
has been measured for GRB~970228 from HST images (Caraveo \etal\ 1997a,b). 
We discuss this issue in some detail.

In addition, although the X-ray, optical and radio afterglows of the GRBs
970228 and 970508 are consistent with the 
cosmological fireball models (Wijers, Rees, and M\'esz\'aros 1997;
Waxman 1997; Vietri 1997; and see \S 2.1), little progress has
been achieved towards the identification of the actual mechanism
causing the bursts. We discuss some of the
implications of the neutron-star merger scenario, and show that the
predicted $\log N - \log P$ relation from this model
is consistent with the observations.

\section{GRB 970508}

The optical counterpart of GRB~970508 was first identified by one of
us (Bond 1997), observing at Kitt Peak National Observatory
(KPNO) on the 0.9-m reflector  with a $2048\times2048$ CCD
camera which provides a field of view of $23'\times23'$.  Upon
notification of the occurrence of the $\gamma$-ray burst by
J.~Halpern, Bond obtained CCD frames beginning at 1997 May 9.131, only
5.5~hours after the burst.  However, as the GRB counterpart was well
below the limits of the STScI Digitized Sky Survey, its actual
identification was delayed until the following night, May 10, when it
was then readily recognized as a variable source by blinking the
frames from both nights. 

Fig.~1 (Plate~1) shows the discovery frames from May 9 and 10. The
variable star-like object at the center of each panel is the proposed GRB
counterpart, which brightened by about 1 mag over the one-day
interval.  The frames, obtained in the standard Johnson $V$ and Kron-Cousins
 $I$ bandpasses, were corrected for atmospheric extinction and
calibrated using 14 standard stars from Landolt (1992).  Table~1
presents the 0.9-m photometry, with $1\sigma$ errors calculated from
the photon statistics using IRAF's {\it qphot\/} routine.  In Fig.~2
we plot the data from Table~1.  The errors are relatively large on
May~9, due both to the faintness of the source and to the relatively
short exposures (120 to 600~s).  On May~10, however, the precision of
the observations is sufficient to reveal brightening of the source
from one frame to the next. 

The principal features of the GRB afterglow at X-ray and optical
wavelengths are well represented by a forward-radiating, blast-wave
model of M\'esz\'aros and Rees (1997) with only four adjustable
parameters, as shown in Fig.~3a.  In their Model a1 those parameters
are:  the peak flux, $F_p$ (which is independent of photon energy),
the continuum power-law photon indices $\alpha$ and $\beta$ for photon
energies less than and greater than, respectively, the peak flux
density, and the duration of the $\gamma$-ray burst, $t_\gamma$.
Values of the parameters which represent the afterglow of GRB 970508
are $F_p = 6\times10^{-5}$~Jy, $\alpha =0.0$, $\beta = -0.9$, and
$t_\gamma = 12~s$ (Fig.~3a).  For this value of $\beta$ the remnant
fades as $t^{-1.4}$.  The values of $\alpha$ and $\beta$ accord well
with those expected for synchrotron emission by relativistic electrons
in a blast wave (M\'esz\'aros and Rees 1997), and the fitted value of
$t_\gamma$ matches well the observed duration of the $\gamma$-ray
burst: 15~s (Costa  \etal\ 1997c), or FWHM = 3.6~s and total
duration = 35~s (Kouveliotou \etal\ 1997). The  $\gamma$-ray and
X-ray brightness of the burst exceed the peak flux density of the
afterglow by a factor $\sim10$ (Fig.~3a). This is probably
a consequence of the fact that the $\gamma$-rays and the afterglow
are produced at different stages of the fireball evolution
(the GRB is produced before the self-similar stage; Waxman 1997a,b).
However, this model represents two general
characteristics of the optical afterglow:  the wavelength dependence
of the fading optical afterglow ($\Delta t > 2$ d), and the absence of
wavelength dependence at earlier times. But the model does not match
(in detail) the temporal evolution before peak brightness.  The
optical brightness before peak can be approximated with
$\alpha = 0.5$, although the optical brightness before peak would then
be predicted to be wavelength dependent, which is contradicted by
observations.  We therefore prefer the $\alpha = 0$ fit. 

A more detailed comparison between the model presented in Fig.~3a and
the optical observations is shown in Fig.~3b.  The ratio of observed
and modeled flux densities are wavelength independent both before and
after peak brightness ($\Delta t \sim 2$ d).  Before peak the observed
flux density rises rapidly from approximately 5 times fainter than
predicted (see however Waxman 1997b) to agreement with the model.  Resolution of this discrepancy
during the early phase of the afterglow may require intensive, rapid
observations of new GRB counterparts.  Generally speaking, given the
simplicity of the model and the small number of adjustable parameters,
the agreement with X-ray, optical and (to some extent) radio observations 
should be considered
remarkable (see also Wijers \etal\ 1997; Vietri 1997; Waxman 1997a,b).
This strengthens considerably the identification of the optical counterparts
of both GRB 970228 and GRB 970508. It is beyond the scope of the
present paper to discuss the complex behavior of GRB 970508 in the
radio wavelengths (Frail et al, 1997).

We have undertaken an extensive analysis of the HST optical images of
GRB~970508 (see also Pian \etal\ 1997), in an attempt to 
detect an unambiguous signature of an  underlying  host galaxy (as
suggested by the line strengths and the line-strength ratios of the Mg~I/Mg~II
absorption system and the [O~II] emission line; Metzger \etal\
1997a,~b).  To this end, we have used the most
updated dark frames; and an average point spread function (PSF)
constructed from the stars in the observed field of the GRB 
(rather than from the archive) which,
within the uncertainties, is indistinguishable from the PSF derived
for the GRB. Our analysis suggests that the limit quoted by Fruchter
\etal\ (1997) is rather conservative,  and we find that if a host
galaxy is underlying the GRB (with an angular extent of about 
0\secpoint5), it must be fainter than $R$ $\sim25.5$ (or else it is extremely
compact with an angular extent of less than about 0\secpoint2, which
at $z = 0.835$ corresponds to $\sim 1.7$~kpc). 
To confirm this result, we did the following exercise: we took the galaxy 
(size $\sim$ 0\secpoint7$\times$0\secpoint9) with r $\sim$ 24.8 at about
5\secpoint6 NE of the GRB (Djorgovski et al. 1997d),
artificially made it fainter by different factors, superposed the 
result on the GRB, and checked whether the galaxy is still detectable.
The galaxy is clearly detected when its magnitude is  $R$ $\sim$ 25.5, but 
barely so at R$\sim$ 26, after the proper PSF subtraction.
Note that this process not only adds the galaxy but also adds the sky noise,
 so even  this method gives a conservative limit to the
detectable magnitude. Since the redshift is   \apgt 0.8, this
implies that an underlying galaxy, if present,  is at least 10 times 
fainter than an L$^*$ galaxy.

\section{On the Proper Motion of GRB~970228}

GRB 970228 was observed with HST Wide Field and Planetary Camera 2
(WFPC2) in `wide-$V$' and `$I$' filters on 1997 March~26 and April~7. 
The telescope roll angles during the two epochs of observation differed by
about 2.\arcdeg3982 (for full details of the observations see Sahu
\etal\ 1997).  On the basis of these observations 
Caraveo \etal\ (1997a,b) reported that the point source  
moved  0\secpoint016 towards the south east over this 12-day
interval. If correct, such a proper motion would imply a Galactic
origin for this GRB. Here we present a detailed independent analysis
of any possible proper motion. 

For the  proper-motion analysis, it is important to correct for
cosmic-ray (CR) and hot-pixel effects in the images.  The standard
STScI pipeline calibration typically uses dark frames taken a few
days prior to the observations, and therefore  may not remove all the
hot pixels efficiently.  Consequently, we have re-calibrated the
images ourselves, starting with the raw, uncalibrated images, and
using the dark frames taken closest (within one day) to the actual
observations.  The bias subtraction, flat-field correction and
CR-rejection were carried out in the usual way. Since there are 
 refractive elements in the optics,  color terms could play a role
in the image positions. Therefore the analysis was done for the
$V$ and $I$ images separately,
giving independent proper-motion information in the two bands. 
Two of the $V$ images had CRs close to the
GRB image, and they were therefore treated with special care. 
Specifically, we carried out two analyses, one with the CR-affected
(and CR-corrected) images included, and one where we simply discarded the 
CR-affected images. (CRs and
their rejection cause loss of information in the affected pixels; thus
discarding such images may be the preferred procedure.) There are
4 reference stars (marked in Fig. 4) in the PC chip whose colors and
brightness imply that their expected parallaxes and proper
motions are orders of magnitude smaller than the
measurement uncertainties.

The centroids of the stars were determined  using  2-dimensional
Gaussian fits. The positional accuracy for the  stars is 
typically  2 to 3 milliarcsec in each coordinate at each epoch, 
while for the relatively faint GRB it is $\sim$4 milliarcsec (corresponding 
to a ``proper-motion" uncertainty of $\sim$6 milliarcsec in each coordinate).  
The measured coordinates were then corrected for the camera's geometric 
distortion using the
Gilmozzi \etal\ (1995) and Holtzman \etal\ (1995) solutions (these
give identical results to within 0.5 milliarcsec, and slightly
differ only at the edge of the field).  A linear
transformation from the first epoch to the second was then performed,
using the 4~reference stars and assuming zero mean motion.  
This was done with 2 different methods: one where the rotation and the
translation were done separately, and the other where they were
done in a single step.
The first method involves using the program `metric' in the
STSDAS package, which first takes the rotation into account (from the
roll-angle information in the header) in converting
the pixel coordinates to RA and DEC. The centroids of the stars are  
determined in RA and DEC, and  a linear translation 
from the first to the second epoch is
then performed as a separate step. In the second method,
the centroids are determined in pixel coordinates taking only
the geometric distortion of the detector into account, and a transformation
from the first to the second epoch is then made taking the rotation and 
translation as a single step.
Both methods, however, gave almost identical results (to within 1 
milliarcsec), which shows that
the telescope roll angle infomation in the header has sufficient
precision for this analysis. The resulting 
``proper motions" for the 4
stars and the GRB are shown in Fig.~5 for V and I filters separately.  
As can be seen,
the positions of the reference stars are within the expected 
uncertainties, which shows that the transformations have
been done correctly. 
The GRB displays a ``proper motion" relative to the reference stars
of about 0\secpoint007  in the NW direction  in the CR-free $V$-band images, 
and  about 0\secpoint0065 in the same direction in the
$I$-band, the uncertainties being of comparable magnitude to the motion. 
The fact that the observed shift is not significantly larger 
than the measurement errors, coupled with the fact 
that the GRB is faint and embedded in a nebulosity,
lead us to  conclude that no proper motion has been detected
(within the uncertainties). Our result 
disagrees with the findings of Caraveo \etal\
(1997), both in magnitude and in the direction of motion, for the $V$-band
images. No comparison is possible for the $I$-band since they
do not quote a value for these images. 

\section{Neutron-Star Mergers}

While  a generic fireball model (\eg \  Paczy\'nski and Rhoads 1993; 
Rees and M\'esz\'aros 1997)
appears consistent with all the available data,  the same cannot yet
be said about any specific mechanism for the production of the
fireball.  In the following we examine several aspects of neutron-star
mergers (see also Sahu \etal\ 1997, and references therein) in 
light of the recent observations and theoretical developments. 

\subsection{The redshift distribution of neutron-star mergers} 

Neutron-star mergers in close-binary systems
have been suggested as a promising mechanism for
the production of fireballs (\eg,  Eichler \etal\ 1989; Narayan, Paczy\'nski, 
and Piran 1992).  Population
synthesis calculations which take into account the evolution of the
entire binary population of a galaxy, (\eg\ Tutukov and Yungelsson 1994)
predict that the frequency of mergers should peak at about $3~\times
10^7$~yr following a burst of star formation. 
Consequently, we can
use recent findings on the cosmic history of star formation to predict
the redshift distribution of neutron-star mergers, and thereby of
GRBs, if such mergers indeed represent the correct model for their
origin. 

Using the Hubble Deep Field (Williams \etal\ 1996) in conjunction with
ground-based observations, Madau \etal\ (1996; see also Madau 1996, 1997)
were able to show that the cosmic star-formation rate SFR(z) peaked at
$z \sim 1.25$. Given the short delay time of $\sim$ 3 $\times$ 10$^7$
yr, we therefore {\it predict} that the rate of neutron-star mergers
should roughly follow the SFR(z) of Madau \etal\ (1997) 
(even delays of up to $\sim 10^9$~yr are hardly noticeable within the
uncertainties). Furthermore, since much of the star formation may be
occurring in small, low-mass galaxies that experience short starbursts
(\eg,  Babul and Rees 1992; Babul and Ferguson 1996), we {\it predict}
that, if GRBs are produced by neutron-star mergers, they may
originate {\it preferentially} in such galaxies. 

We have checked whether this strongly non-uniform redshift distribution
(Madau 1997) is compatible with the observed $\log N(>P) - \log P$
relationship for GRBs (where $N(>P)$ is the number of bursts with peak
flux greater than P[photons cm$^{-2}$ s$^{-1}$]).  We have used a
standard candle luminosity distribution function for a range of
luminosities $L_\gamma = 10^{49} - 10^{53}$ erg$^{-1}$~s$^{-1}$ (100 - 500 keV),
a power-law photon spectrum with index $-1.5$, and Friedmann cosmology
with $q_0 = 0.2$ and $H_0$ = 50~\mh.  The synthetic $\log N - \log P$
distribution functions thus obtained are presented in Fig.~6.  Also
presented in Fig.~6 is the distribution function observed with the
Pioneer Venus Orbiter (PVO) and the Compton Gamma-Ray Observatory
Burst and Transient Source Experiment (BATSE) (Fenimore \etal\ 1993, Fenimore and
Bloom 1995).  A reasonable representation of the observed distribution
is found for $L_{\gamma} = 10^{51}$~erg$^{-1}~s^{-1}$. Thus 
density evolution that follows the Madau \etal\ SFR function 
and a single peak GRB luminosity are consistent with the
observations. We would like to stress that if neutron-star mergers are
the correct model for GRBs, then their redshift distribution can 
provide {\it an independent test} for the cosmic history of the SFR. 

\subsection{``Kicks"}

Another issue that should be addressed in relation to neutron-star
mergers is that of ``kicks" that are required for the formation of the
neutron-star binaries as inferred, for example, from observations of
radio pulsars (\eg,  Lyne and Lorimer 1994) and low-mass
X-ray binaries (White and van Paradijs, 1996).  In particular, Fryer and
Kalogera (1997) find that the formation rate of double neutron-star
systems with separations that are smaller than $5R_{\odot}$ (a
fraction of which are the candidates for mergers within a Hubble time)
peaks at kick velocities of 200~\kms.  Furthermore, Fryer and Kalogera
find {\it minimum} center-of-mass velocities of 200~\kms\ and 
225~\kms\ for the two double neutron-star systems PSR~1913+16 and 1534+12
(respectively).  %This raises the following interesting possibility:
Since the escape velocity from small galaxies is of order 100~\kms\ 
(\eg,  Gallagher, Hunter, and Tutukov\ 1984), neutron-star binaries
with center-of-mass velocities \apgt 200~\kms\ would easily escape from
such host galaxies. The HST observations of GRB~970228
(Sahu \etal\ 1997) are consistent with such an
interpretation, since the observed point source lies about 0\secpoint3
south of the center of the extended source (which may be the host
galaxy). 

No host galaxy has been detected yet for GRB 970508 (see \S 2).  The HST image,
however, shows two nearby faint galaxies, at transverse distances of $\sim
30$~kpc and 35~kpc (for $z = 0.835$, $H_0 =$ 65~\mh) from the point
source.  A neutron-star binary moving at $\sim 300$~\kms\ (away from a
low-mass galaxy) for about $10^8$~yr could reach a distance of $\sim
32$~kpc.  However, the ratio of the Mg~I/Mg~II lines and the strength
of these lines, and in
particular the reported [O~II] emission line (Metzger \etal\
1997a,~b)  argue for the presence of a host galaxy at the GRB
location. We conclude that if a host is not found (even after the GRB
fades away), an interpretation associating the GRB with one of the
faint galaxies will have to be re-examined.  In such a case, one would
probably have to argue that the [O~II] emission is produced by a
nebula in the vicinity of the GRB itself, in a similar manner 
to the production of emission lines from the ring around SN~1987A
(\eg,  Sonneborn \etal\ 1997).  One test for such an
interpretation would be to spectroscopically monitor GRB~970508, since
the strength of the line would then be expected to change, as in the case of
SN~1987A. 

\acknowledgments
We are grateful to P.~Madau and E.~Fenimore for useful discussions. 
We acknowledge the help of Colin Cox with the geometric distortion correction
of the HST images, and Ray Lucas and others in the scheduling division for 
efficiently scheduling the observations of GRB 970228. 
We thank F.~Frontera and M.~Garcia for communication
of results in advance of publication. ML acknowledges support from
NASA grant NAGW-2678. KPNO is operated by the 
Association of Universities for Research 
in Astronomy, Inc., under contract with the National Science Foundation.

\clearpage

\figcaption[fig1.ps]{ (Plate 1) Discovery images for the optical
counterpart of GRB~970508.  These are coadded $V$-band frames, taken at mean
times of 1997 May 9.19 (top, 600s exposure) and May 10.18 (bottom;
1800s exposure).  Each frame is $138''$ high with north at the top
and east on the left. 
The GRB counterpart is the variable source marked in the bottom frame. 
It brightened by 1~mag between May 9 and 10. \label{fig1}} 

\figcaption[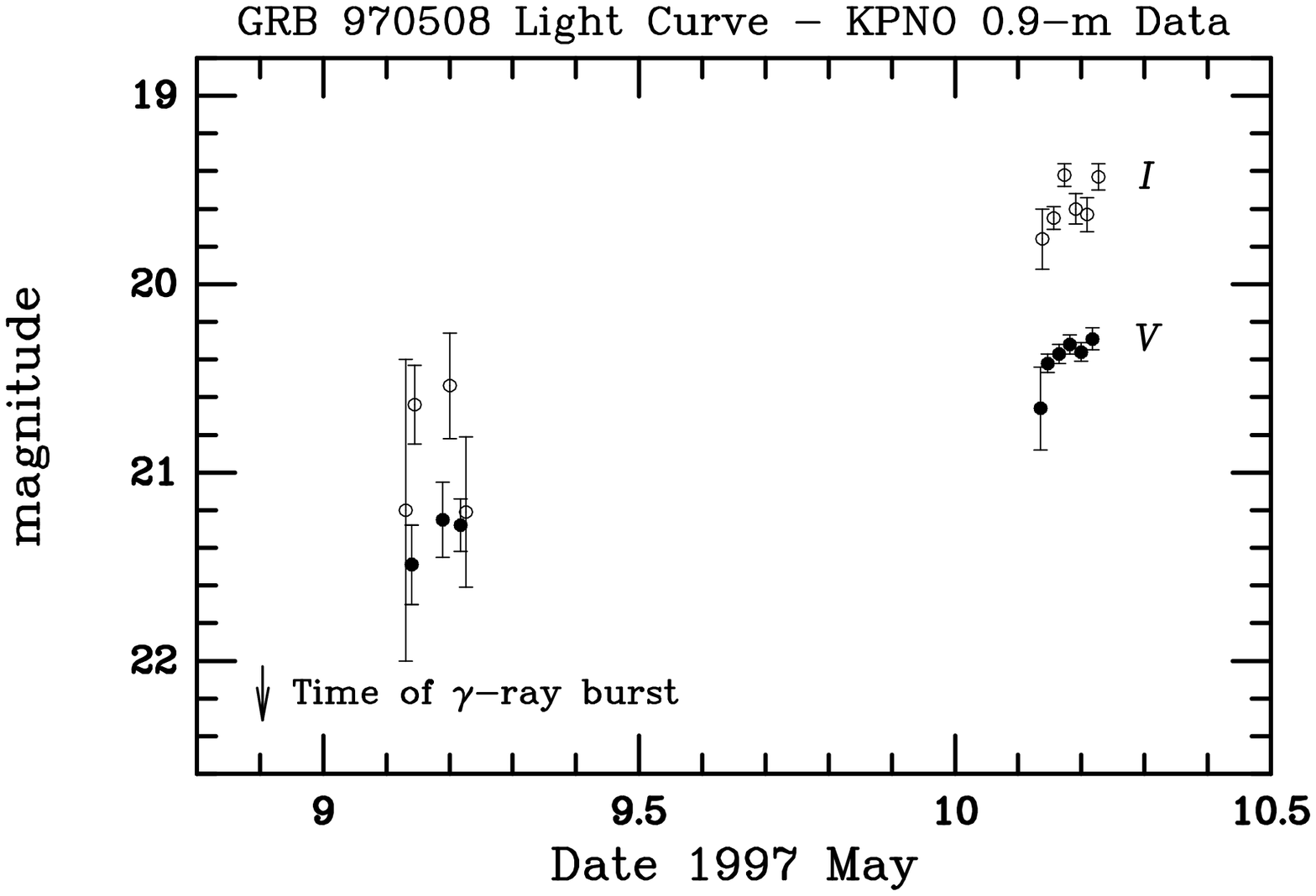]{ Light curve of the optical counterpart of
GRB~970508 from our Kitt Peak observations.  Exposure times ranged
from 120 to 600~s, leading to a range of error bars.  On May 10, the
source was seen to brighten from one frame to the next.
\label{fig2}} 

\figcaption[fig3.ps]{ {\bf (a)} Multi-wavelength observations and
model of the GRB~970508 afterglow. The observed optical flux
densities (Castro-Tirado \etal\ 1997; Chevalier and Ilovaisky 1997;
Djorgovski \etal\ 1997a,~b,~c,~d; Donahue \etal\ 1997; Fruchter \etal\
1997; Galama \etal\ 1997b,~c,~d; Garcia 1997; Garcia \etal\ 1997;
Groot \etal\ 1997; Jaunsen \etal\ 1997; Kopylov \etal\ 1997a,~b;
Metzger \etal\ 1997a; Mignoli \etal\ 1997; Morris \etal\ 1997;
Schaefer \etal\ 1997; and Table 1) are corrected for foreground
Galactic extinction ($E_{B-V} = 0.07$, Burstein and Heiles 1982).  For
clarity only the modeled X-ray, $U, I,$ and $K_s$ light curves are
presented. The $B, V, R,$  and $r$ light curves are intermediate to
those in $U$ and $I$. Model a1 of M\'esz\'aros and Rees (1997; Wijers
\etal\ 1997), with parameter values as described in the text, is
shown.  For illustration, the GRB peak flux densities in $\gamma$-rays
and X-rays are plotted at the measured $\gamma$-ray BeppoSAX GRBM
(Costa \etal\ 1997c) and BATSE (Kouveliotou \etal\ 1997) burst
duration. {\bf (b)} Ratio of the observed and modeled optical flux
density of the afterglow of GRB~970508. \label{fig3a,b}} 

\figcaption[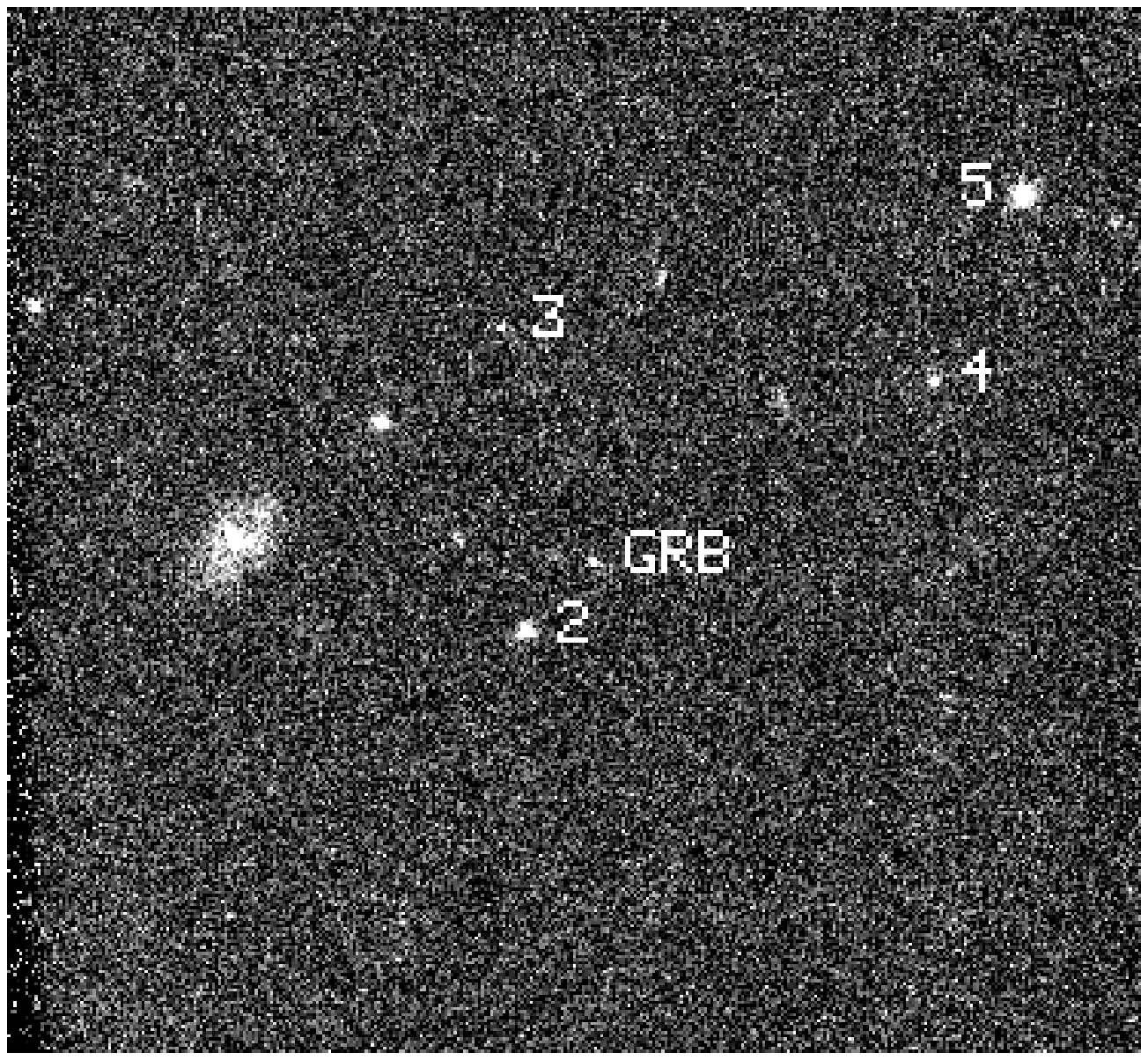]{ (Plate 2) The reference stars used for
the proper-motion analysis of the GRB.
\label{fig4}} 

\figcaption[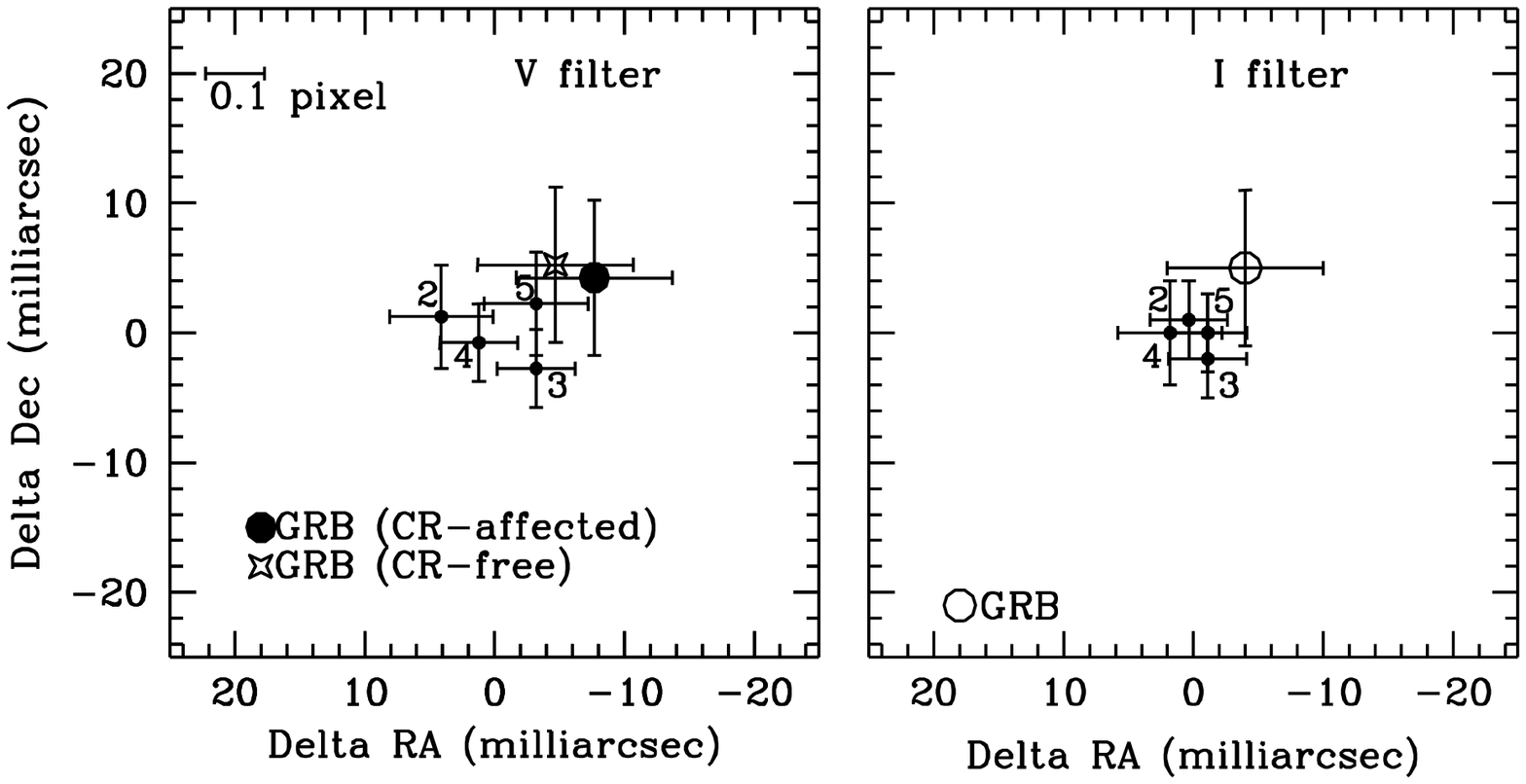]{ Measured ``proper motion"  of the optical 
counterpart of GRB~970228 and the 4 stars in the
PC-chip (see Fig. 4), assuming zero mean motion for the 4 stars. 
The left and the right panels show the results obtained from the
analysis of the V and I images, respectively.
The error bars refer to $\pm 1 \sigma$.
Note that considering only the CR-free images reduces the ``proper 
motion" (see text).
\label{fig5}} 

\figcaption[fig6.ps]{ Representation of GRB source counts with density
evolution of standard candles in a Friedmann cosmology. \label{fig6}} 

\begin{table*}
\begin{flushleft}
{\caption[]{Photometry of GRB~970508 during the first two days after outburst}}
\begin{tabular}{rlcrccccl}
\\
\hline
\\
1997 UT Date&           $V$ mag&  error&    1997 UT Date&      $I$ mag&  error&\\

\hline
\\
&&&                                 May  9.131&   21.2&   0.8&\\
     May  9.140&   21.49&  0.21&          9.145&   20.64&     0.21&\\
    9.189&   21.25&     0.20&          9.200&   20.54&     0.28&\\
     9.217&   21.28&     0.14&          9.226&   21.21&     0.40&\\
     10.135&   20.66&     0.22&         10.138 &  19.76&     0.16&\\
     10.147&   20.42    & 0.05&         10.156  & 19.65&    0.06&\\
     10.165 &  20.37   &  0.05 &        10.173&   19.42 &    0.06&\\
     10.182  & 20.32  &   0.05  &       10.191  & 19.60  &   0.08&\\
     10.200   &20.36 &    0.05   &      10.209 &  19.63   &  0.09&\\
     10.218 &  20.29&     0.06    &     10.227&   19.43    & 0.07&\\
\\
\hline
\end{tabular}
\end{flushleft}

\end{table*}

\end{document}